\documentclass[a4paper,UKenglish,cleveref, autoref, thm-restate]{lipics-v2021}

\newif\ifdoubleblind
\newif\ifcomments
\newif\ifextfig
\newif\iftr

\trtrue

\title{Deep Multilevel Graph Partitioning}

\ifdoubleblind
\author{Anonymous Authors}{Anonymous affiliation}{}{}{}
\authorrunning{A. Authors}
\Copyright{Anonymous Authors}
\else
\author{Lars Gottesbüren}{Karlsruhe Institute of Technology, Germany}{}{}{}
\author{Tobias Heuer}{Karlsruhe Institute of Technology, Germany}{}{}{}
\author{Peter Sanders}{Karlsruhe Institute of Technology, Germany}{}{}{}
\author{Christian Schulz}{Heidelberg University, Germany}{}{}{}
\author{Daniel Seemaier}{Karlsruhe Institute of Technology, Germany}{}{}{}
\authorrunning{L. Gottesbüren and T. Heuer and P. Sanders and C. Schulz and D. Seemaier}
\Copyright{Lars Gottesbüren and Tobias Heuer and Peter Sanders and Christian Schulz and Daniel Seemaier}
\fi
\ccsdesc[500]{Mathematics of computing~Graph algorithms}
\keywords{graph partitioning, graph algorithms, multilevel, shared-memory, parallel}
\ifdoubleblind
\supplement{Link to source code, experimental results and benchmark sets are available, but not provided due to
double-blind review process.}
\else
\supplement{The source code and data has been made available at \url{https://algo2.iti.kit.edu/seemaier/deep_mgp/}.}
\fi

\ifdoubleblind
\else
\acknowledgements{The authors acknowledge support by the State of Baden-Württemberg through bwHPC. This project has received funding from the European Research Council (ERC) under the European Union’s Horizon 2020 research and innovation programme (grant agreement No. 882500).
\iftr
\\ \includegraphics[width=3cm]{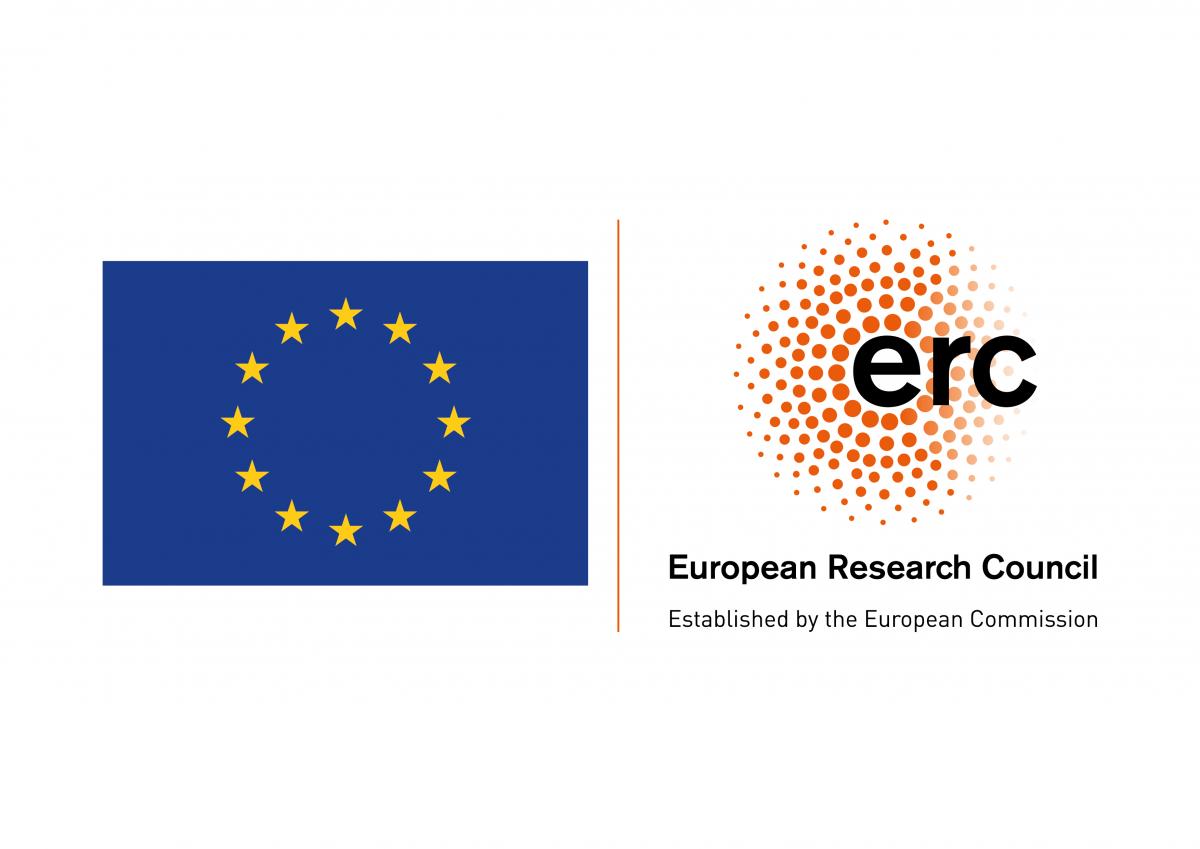} \\ \textcolor{white}{.} 
\fi
}
\fi

\ifdoubleblind
\else
\nolinenumbers
\fi
\iftr
\hideLIPIcs
\fi

\EventEditors{John Q. Open and Joan R. Access}
\EventNoEds{2}
\EventLongTitle{42nd Conference on Very Important Topics (CVIT 2016)}
\EventShortTitle{CVIT 2016}
\EventAcronym{CVIT}
\EventYear{2016}
\EventDate{December 24--27, 2016}
\EventLocation{Little Whinging, United Kingdom}
\EventLogo{}
\SeriesVolume{42}
\ArticleNo{23}

\usepackage{booktabs}
\usepackage[autolanguage]{numprint}
\usepackage{mathtools}
\usepackage{pdflscape}
\usepackage{afterpage}
\usepackage{graphics}
\usepackage{placeins}
\usepackage{cancel}
\usepackage{amsfonts}

\newcommand{\maxdeg}[1]{\ensuremath{\Delta_{#1}}}

\newcommand{\meddeg}{\ensuremath{\widetilde{d(v)}}}

\newcommand{\realrange}[2]{\left[#1, #2\right]}

\newcommand{\unitrange}[2]{\realrange{0}{1}}

\newcommand{\Oh}[1]{\mathcal{O}\!\left( #1\right)}

\newcommand{\Th}[1]{\Theta\!\left( #1\right)}

\newcommand{\llabel}[1]{\label{\labelprefix:#1}}
\newcommand{\labelprefix}{} 

\newcommand{\discussionsize}{\small}

\newcommand{\frage}[1]{}

\newcommand{\myparagraph}[1]{\vspace{-3mm}\subparagraph*{#1}}
\newcommand{\punkt}{\enspace .}

\newenvironment{code}{\noindent
\begin{tabbing}%
\hspace{2em}\=\hspace{2em}\=\hspace{2em}\=\hspace{2em}\=\hspace{2em}\=%
\hspace{2em}\=\hspace{2em}\=\hspace{2em}\=\hspace{2em}\=\hspace{2em}\=%
\kill}{\end{tabbing}}

\newcommand{\labelcommand}{}
\newcommand{\captiontext}{}
\newsavebox{\codeparam}
\newcounter{lineNumber}
\newenvironment{disscodepos}[3]{%
\renewcommand{\labelcommand}{#2}%
\renewcommand{\captiontext}{#3}%
\sbox{\codeparam}{\parbox{\textwidth}{#3}}%
\begin{figure}[#1]\begin{center}\begin{code}\setcounter{lineNumber}{1}}{%
\end{code}\end{center}\caption{\llabel{\labelcommand}\captiontext}\end{figure}}

{\end{disscodepos}}

\newcommand{\While}    {{\bf while\ }}

\newcommand{\Return}   {{\bf return\ }}

\newdimen\endofsize\endofsize=0.5em
\def\endofbeweis{~\quad\hglue\hsize minus\hsize
                 \hbox{\vrule height \endofsize width
\endofsize}\par}

\newcommand{\Partitioner}[1]{\textsf{#1}}
\newcommand{\subgraph}[1]{\ensuremath{G[#1]}}
\newcommand{\cut}[1]{\ensuremath{\text{\texttt{cut$(#1)$}}}}

\numberwithin{equation}{section}
\usepackage{siunitx}
\usepackage{wasysym,amssymb,pifont}
\usepackage{thm-restate}
\usepackage{pgfplots}
\usepgfplotslibrary{external}
\tikzsetexternalprefix{tr_plots/}

\usepackage{marvosym}

\usepackage[algo2e,ruled,vlined,linesnumbered]{algorithm2e}
\SetCommentSty{textit}
\DontPrintSemicolon
\IncMargin{-\parindent}
\SetAlCapHSkip{0pt}
\SetAlgoLined
\makeatletter
\newcommand{\remotelatexerror}{\let\@latex@error\@gobble}
\makeatother

\usepackage{wrapfig}

\makeatletter
\def\maxwidth{ %
  \ifdim\Gin@nat@width>\linewidth
  \linewidth
  \else
  \Gin@nat@width
  \fi
}
\makeatother

\usepackage{alltt}
\newcommand{\ie}{i.e.}
\newcommand{\etal}{et~al.}

\usepackage[square,numbers,sort&compress]{natbib}
\usepackage{float}

\definecolor{fuchsiapink}{rgb}{1.0, 0.47, 1.0}

\ifcomments
\newcommand{\psan}[1]{{\color{red}[PS: #1]}}
\newcommand{\csch}[1]{{\color{blue}[CS: #1]}}
\newcommand{\dsee}[1]{{\color{black!50!green}[DS: #1]}}
\newcommand{\lars}[1]{{\color{brown}[LG: #1]}}
\newcommand{\larscancel}[1]{{\color{brown}{\cancel{#1}}}}
\newcommand{\tobi}[1]{{\color{fuchsiapink}[TH: #1]}}
\else
\newcommand{\psan}[1]{{}}
\newcommand{\csch}[1]{{}}
\newcommand{\dsee}[1]{{}}
\newcommand{\lars}[1]{{}}
\newcommand{\larscancel}[1]{{}}
\newcommand{\tobi}[1]{{}}
\fi

\usepackage{dcolumn}
\newcolumntype{d}[1]{D{.}{.}{#1}}

\begin{document}
  \maketitle

  \begin{abstract}
Partitioning a graph into blocks of ``roughly equal`` weight while cutting only few edges is a fundamental problem in
computer science with a wide range of applications.
In particular, the problem is a building block in applications that require parallel processing.
While the amount of available cores in parallel architectures has significantly increased in recent years,
state-of-the-art graph partitioning algorithms do not work well if the input needs to be partitioned into a large number
of blocks.
Often currently available algorithms compute highly imbalanced solutions, solutions of low quality, or have excessive
running time for this case.
This is due to the fact that most high-quality general-purpose graph partitioners are \emph{multilevel algorithms} which
perform graph coarsening to build a hierarchy of graphs, initial partitioning to compute an initial solution, and local
improvement to improve the solution throughout the hierarchy.
However, for large number of blocks, the smallest graph in the hierarchy that is used for initial partitioning still has
to be large.

In this work, we substantially mitigate these problems by introducing \emph{deep multilevel graph partitioning} and a
shared-memory implementation thereof.
Our scheme continues the multilevel approach deep into initial partitioning -- integrating it into a framework where
recursive bipartitioning and direct $k$-way partitioning are combined such that they can operate with high performance
and quality.
Our integrated approach is stronger, more flexible, arguably more elegant, and reduces bottlenecks for parallelization
compared to existing multilevel approaches.
For example, for large number of blocks our algorithm is on average at least an order of magnitude faster than competing algorithms while
computing partitions with comparable solution quality.
At the same time, our algorithm consistently produces balanced solutions.
Moreover, for small number of blocks, our algorithms are the fastest among competing systems with comparable quality.
  \end{abstract}

  \vfill\pagebreak

  \clearpage
  \setcounter{page}{1}
\section{Introduction}\label{sec:introduction}
\vspace{-2mm}Graphs are a universal abstraction for modelling relations between
objects.  Thus they are used throughout computer science and have
applications with an ever growing volume and variety of the considered
graphs.  One frequently needed basic operation is \emph{balanced graph
  partitioning} -- cutting a graph into $k$ pieces of ``roughly
equal'' weight while cutting only few edges.
Balanced graph partitioning is NP-hard and
even NP-hard to approximate~\cite{andreev2006balanced} and thus
usually solved using heuristics.  In particular, \emph{multilevel
  graph partitioning (MGP)} is used in most high-quality
general-purpose systems: During \emph{coarsening}, build a hierarchy
of graphs where each graph is a coarse approximation of the previous
one. When the coarse graph is ``small'', run a possibly expensive
\emph{inital partitioning} method on it.
This is useful because a feasible partition at the coarsest level is a feasible partition of
the original input with the same cut value.
The partition is successively \emph{uncoarsened} to each finer level
and \emph{locally improved}. This is often both faster and higher quality than applying
comparable improvement algorithms only on the finest level since MGPs have a more global view on the coarse levels and can move
entire groups of nodes in constant time.

A prominent application (out of many) is distributing workload across parallel machines with little communication.
With growing numbers of processors in parallel machines, we are interested in large values of $k$ -- in the order of millions.
However, existing research has mostly focused on small values, typically $2 \leq k \leq 256$.
Unsurprisingly, these systems perform poorly for large $k$.
If \emph{direct $k$-way partitioning} is used, the coarsest graph still has to be large when initial partitioning is called.
\emph{Recursive bipartitioning} performs $\log(k)$ cycles of (un)coarsening and is either restricted to very small imbalances or unlikely to return a feasible solution.
Further, both exhibit parallelization bottlenecks on coarse levels.

\begin{figure}
    \centering
    \includegraphics[width=\textwidth]{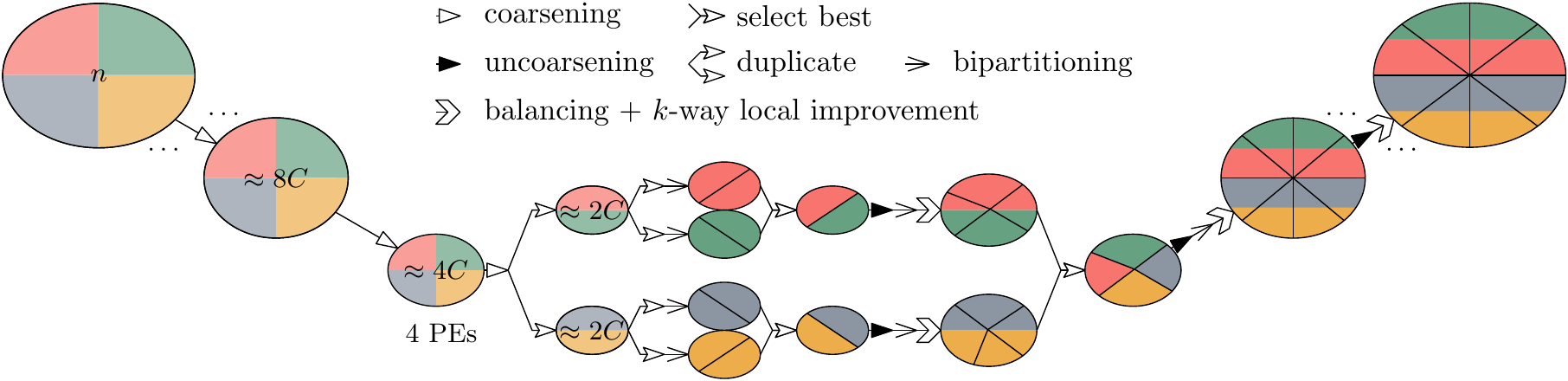}
    \caption{Outline of deep MGP. Duplicate while coarsening to maintain load $\geq C$ on each PE.
    Successively bipartition during uncoarsening.
    Colors indicate work performed by each PE.}
    \label{fig:part_scheme}
  \end{figure}

We mitigate these problems by introducing \emph{deep MGP}, an approach that continues the multilevel scheme deep into initial
partitioning and integrates aspects of direct $k$-way and recursive bipartitioning.
Deep MGP can be instantiated with concrete (parallel) building blocks for (un)coarsening, $k$-way local improvement, and bipartitioning of small graphs.
We also include \emph{balancing} as an explicit component.
Figure~\ref{fig:part_scheme} summarizes the approach.
Deep MGP performs only one cycle of (un)coarsening.
Bipartitioning is done during uncoarsening so that it is always applied to graphs with about $C$ nodes (input parameter) until
the desired number of $k$ blocks is reached.
To exploit all the available
parallelism, the invariant is maintained that parallel tasks performed
by $x$ processing elements (\emph{PEs}) work on graphs with at least
$xC$ nodes. Maintaining this invariant during coarsening allows multiple diversified attempts with little overhead invested.

Under certain simplifying assumptions, deep MGP for $k$-partitioning an
$n$-node graph with bounded degree can be done in time $\Oh{(n/p)\max(1,\log(kC/n))+\log^2 n}$, i.e.,
with linear work and polylogarithmic span unless $k$ is very large; see \cref{sec:deep_mgp}.

After introducing notation and basic concepts in
\Cref{sec:preliminaries} and discussing related work in
\Cref{sec:relwork}, \Cref{sec:deep_mgp} introduces deep MGP as a
generic method.
In \cref{sec:instantiation}
we describe the simple and fast \Partitioner{KaMinPar} partitioner
which uses deep
MGP to achieve scalability to both large $k$ and a considerable
number of parallel cores while guaranteeing balanced solutions.

  In \Cref{sec:eval} we report on extensive experiments which indicate that \Partitioner{KaMinPar} has a very favorable
  quality-performance tradeoff and very good scaling behavior.
  For traditional values of $k$, it is faster than previous algorithms that can achieve comparable or better quality.
  For large $k$, previous algorithms mostly find infeasible solutions and exhibit excessive running times, whereas \Partitioner{KaMinPar} consistently finds feasible solutions with comparable quality and is an order of magnitude faster.

  \Cref{sec:conclusion} summarizes the results and outlines possible future directions.

  \section{Preliminaries}\label{sec:preliminaries}
\vspace{1mm}
  \myparagraph{Notation and Definitions.}

  Let $G = (V, E, c, \omega)$ be an undirected graph with node weights
  $c: V \rightarrow \mathbb{N}_{> 0}$, edge weights $\omega: E \rightarrow \mathbb{N}_{> 0}$,
  $n \coloneqq \lvert V \rvert$, and $m \coloneqq \lvert E \rvert$.
  We extend $c$ and $\omega$ to sets, \ie, $c(V') \coloneqq \sum_{v \in V'} c(v)$ and
  $\omega(E') \coloneqq \sum_{e \in E'} \omega(e)$. $N(v) \coloneqq \{ u \mid \{u, v\} \in E\}$
  denotes the neighbors of $v$ and $E(v) \coloneqq \{ e \mid v \in e\}$ denotes the edges incident to $v$.
  For some $V' \subseteq V$, $\subgraph{V'}$ denotes the subgraph of $G$ induced by $V'$.
  We are looking for \emph{blocks} of nodes $\Pi \coloneqq \{ V_1, \dots, V_k \}$ that partition $V$, \ie, $V_1 \cup \dots \cup V_k = V$ and
  $V_i \cap V_j = \emptyset$ for $i \neq j$.
  The \emph{balance constraint} demands that
  $\forall i \in \{1..k\}: c(V_i) \le L_{\max,k} \coloneqq \max\{(1 + \varepsilon) \frac{c(V)}{k}, \frac{c(V)}{k} + \max_v c(v)\}$
  for some imbalance parameter $\varepsilon$.\footnote{Traditionally,
  $L_{k} \coloneqq (1 + \varepsilon)\lceil \frac{c(V)}{k} \rceil$ is used as balance constraint.
  However, finding a balanced partition with $L_k$ is NP-complete, which, as we will
  see in Section~\ref{sec:deep_mgp}, is not case for $L_{\max,k}$.}
  The objective is to minimize $\cut{\Pi} \coloneqq \sum_{i < j} \omega(E_{ij})$ (weight of all cut edges), where
  $E_{ij} \coloneqq \{\{u, v\} \in E \mid u \in V_i, v \in V_j \}$.
  We call a node $v \in V_i$ that has a neighbor $w \in V_j$, $i \neq j$ a \emph{boundary node}.
  A \emph{clustering} $\mathcal{C} \coloneqq \{C_1, \ldots, C_l\}$ is also a partition of $V$, where
  the number of blocks $l$ is not given in advance (there is also no balance constraint).

  \myparagraph{Multilevel Graph Partitioning.}
  Many high-quality graph partitioners employ the multilevel paradigm,
  which consists of three phases:
  During the \emph{coarsening phase}, the algorithms build a hierarchy of successively smaller graphs
  where each graph is a coarse approximation of the previous one.
  Coarse graphs are built by either computing node clusters or matchings and afterwards
  \emph{contracting} them.
  A clustering $\mathcal{C} = \{C_1, \ldots, C_l\}$ is contracted by collapsing each cluster $C_i$ into a single node $c_i$
  with weight $c(c_i) = \sum_{v \in C_i} c(v)$.
  There is an edge $e = (c_i, c_j)$ in the contracted graph with weight $\omega(e) = \sum_{(u,v) \in E_{ij}} \omega(u,v)$
  where $E_{ij}$ are the edges that connect cluster $C_i$ and $C_j$ in the original graph, if $|E_{ij}| > 0$.
  Once the number of nodes of a coarse graph falls below a certain threshold
  or the coarsening algorithm converges,
  \emph{initial partitioning} computes a partition of the coarsest graph.
  Finally, \emph{refinement} subsequently undoes the contractions performed during coarsening.
  In each uncontraction, the partition is first projected to the finer graph and then improved using \emph{local improvement}
  algorithms.

  Generally, there are two ways to partition a graph into $k$ blocks using the mutlilevel paradigm, namely
  \emph{direct $k$-way} partitioning and \emph{recursive bipartitioning}.
  The former coarsens the graph until $\Omega(k)$ nodes are left -- usually $kC$ nodes where $C$ is an input parameter --
  and then computes a $k$-way partition of the coarsest graph.
  The latter first computes a bipartition $\Pi = \{V_1, V_2\}$ and then recurses on the induced subgraphs $\subgraph{V_1}$ and
  $\subgraph{V_2}$ by partitioning $V_1$ into $\lceil \frac{k}{2} \rceil$ and $V_2$ into $\lfloor \frac{k}{2} \rfloor$ blocks.
  Note that many multilevel graph partitioners based on the direct $k$-way partitioning scheme use recursive bipartitioning
  to compute an initial partition of the coarsest graph~\cite{KarypisK98a, kaffpa, lasalle2013multi, ahss2017alenex, MT-KAHYPAR-FAST}.

  \myparagraph{Size-Constrained Label Propagation.}
  Based on the \emph{label propagation clustering} algorithm by Raghavan~\etal~\cite{labelpropagationclustering},
  Meyerhenke~\etal~\cite{LPAgraphPartitioning} introduced the \emph{size-constrained label propagation} algorithm as a
  coarsening and refinement algorithm.
  The algorithm is parameterized by a maximum cluster size $U$.
  In the coarsening resp.~refinement phase, each node is initially assigned to its own cluster resp.~to its corresponding block of the partition.
  The algorithm then works in rounds.
  In each round, the nodes are visited in some order and a node $u$ is
  moved to the cluster resp.~block $K$ that contains the most neighbors of $u$ without violating the size constraint,
  \ie, $c(K) + c(u) \le U$.
  The algorithm terminates once no more nodes were moved during a round or a maximum number of rounds has been
  exceeded.

  \myparagraph{Maintaining the Balance Constraint.}
  Finding a balanced partition of a weighted graph with $L_k \coloneqq (1 + \varepsilon) \lceil \frac{c(V)}{k} \rceil$
  as balance constraint is NP-complete as it
  can be reduced to the problem of scheduling jobs on identical parallel
  machines~\cite{garey1979computers}. Therefore, many partitioners
  incorporate techniques that prevent the formation of heavy nodes
  during the coarsening process by penalizing the contraction of nodes
  with large weights~\cite{catalyurek1999hypergraph} or enforcing a
  strict upper bound for node weights~\cite{hs2017sea}.
  This makes it easier for initial partitioning to find a feasible initial
  solution. However, as we will see in Section~\ref{sec:deep_mgp}, if we replace $L_k$
  with $L_{\max,k}$ the problem of finding a balanced partition becomes solvable in
  polynomial time.
  In the recursive bipartitioning setting, using the input imbalance parameter $\varepsilon$ for each
  bipartition can produce blocks in the final $k$-way partition that would violate the balance constraint.
  Therefore, \Partitioner{KaHyPar}~\cite{KAHYPAR-BP,KaHyPar-R} ensures
  that a $k$-way partition obtained via recursive bipartitioning is
  balanced by adapting the imbalance ratio for each
  bipartition individually. Let $\subgraph{V'}$ be the
  subgraph of the current bipartition that should be partitioned
  recursively into $k' \le k$ blocks. Then,
  $\scriptstyle \varepsilon' \coloneqq \left( (1 + \varepsilon) \frac{c(V)}{k} \cdot \frac{k'}{c(V')} \right)^{\frac{1}{\lceil \log_2(k') \rceil}}~-~1$
  is the imbalance ratio used for the bipartition of $\subgraph{V'}$. If each bipartition is $\varepsilon'$-balanced, then the final $k$-way partition
  is $\varepsilon$-balanced.

  \section{Related Work}\label{sec:relwork}

  There has been a huge amount of research on graph partitioning so that we refer the reader
  to overview papers~\cite{schloegel2000gph,GPOverviewBook,Walshaw07,buluc2016recent,DBLP:reference/bdt/0003S19} for most of the
  material and to Appendix~\ref{appendix:parallel_graph_partitioning} for a detailed overview of techniques used in parallel
  multilevel graph partitioners.
  Here, we focus on issues closely related to our main contributions.
  Most modern high-quality graph partitioners are based on the multilevel paradigm.
  Well-known software packages based on this approach include \Partitioner{KaHIP}~\cite{kaffpa} and
  \Partitioner{Metis}~\cite{karypis1998fast} (sequential graph partitioners),
  \Partitioner{Mt-KaHIP}~\cite{DBLP:journals/tpds/AkhremtsevSS20} and \Partitioner{Mt-Metis}~\cite{lasalle2013multi,lasalle2016parallel} (shared-memory
  graph partitioners), \Partitioner{Jostle}~\cite{Walshaw07}. \Partitioner{PT-Scotch}~\cite{ptscotch}, \Partitioner{ParHIP}~\cite{DBLP:journals/tpds/MeyerhenkeSS17} and
  \Partitioner{ParMetis}~\cite{karypis1996parallel} (distributed graph partitioners).

  \Partitioner{Mt-KaHIP}~\cite{DBLP:journals/tpds/AkhremtsevSS20} uses size-constrained label propagation in the coarsening and refinement phase.
  Additionally, it implements a parallel direct $k$-way FM algorithm based on the \emph{localized multi-try}
  FM used in \Partitioner{KaHIP}~\cite{kaffpa}. In the initial partitioning phase, it performs independent
  runs on each PE using \Partitioner{KaHIP} and apply the partition with the smallest edge cut to the coarsest graph.

  \Partitioner{Mt-Metis}~\cite{lasalle2013multi,lasalle2016parallel} uses matching-based coarsening and
  parallel recursive bipartitioning to compute an initial partition of the coarsest graph. For refinement, it implements
  a $k$-way greedy refinement (FM with only positive gain moves), and hill-scanning algorithm,
  a simplification of localized FM where small groups of vertices, whose individual gains are negative,
  are moved if their combined gain is positive.

\myparagraph{Limitations of Existing Systems for Large $k$.}
The coarsening phase of MGP usually stops when $kC$ nodes are left.
For large $k$, this breaks the assumption that the coarsest graph is small.
Thus, really expensive initial partitioners are infeasible at this level.
Many MGPs therefore use multilevel recursive bipartitioning for the coarsest graph~\cite{KarypisK98a, kaffpa, lasalle2013multi, ahss2017alenex, MT-KAHYPAR-FAST}.
This results in a sequential running time of $\mathcal{O}(T\log{k})$ where $T$ is the
running time of the bipartitioning algorithm. When this is used within a parallel algorithm,
initial partitioning can become a major bottleneck~\cite{DBLP:journals/tpds/AkhremtsevSS20}.

Furthermore, a feasible solution for the $k$-way graph partitioning problem must satisfy
the balance constraint that usually depends on the average block weight $\frac{c(V)}{k}$.
Thus, increasing the number of blocks leads to a tighter balance constraint and
drastically reduces the space of feasible solutions.
Therefore, a partitioner
handling larger values of $k$ has to employ techniques to ensure that the
balance constraint is not violated.

  \section{Generic Deep MGP}\label{sec:deep_mgp}

  In this section, we present our first major contribution -- a new partitioning scheme that we call
  \emph{Deep Multilevel Graph Partitioning}.
  Deep MGP continues coarsening deep into the initial partitioning phase.
  Roughly speaking, it starts by coarsening the input graph until $2C$ nodes are left -- for some input parameter $C$.
  Then, the coarsest graph is bipartitioned into two blocks.
  During uncoarsening, it maintains the key invariant that the partition of a coarse graph with~$n'$ nodes has
  $k' = \min\{k, \textrm{ceil}_2(\frac{n'}{C}) \}$ blocks, where $\textrm{ceil}_2(x)$ is $x$ rounded up to the next power
  of two.
  This value is chosen such that each invocation of flat bipartitioning works on a graph with roughly $2C$ nodes.
  Thus, the graph is divided into $\min\{k, \textrm{ceil}_2(\frac{n}{C})\}$ blocks after unrolling the graph hierarchy.
  If $k > \textrm{ceil}_2(\frac{n}{C})$, blocks are further subdivided until $k$ blocks are obtained.

  This approach combines the merits of direct $k$-way partitioning and
  recursive bipartitioning: similar to direct $k$-way partitioning, deep MGP
  coarsens and uncoarsens the graph only once, and enables the use of
  $k$-way local improvement algorithms throughout the graph
  hierarchy.
  Moreover, it enforces that (possibly expensive) bipartitioning algorithms
  are only applied to small graphs.
  Thereby, it eliminates the initial partitioning bottleneck for large
  values of $k$ or parallel graph partitioning.

  In the remainder of this section, we give a detailed description of deep MGP.
  We simplify this description by restricting $k$ to powers of two, but lift this restriction in a subsequent paragraph.
  Finally, we describe how to parallelize it and analyze its running time.

  \begin{figure}
    \begin{minipage}[t]{0.48\textwidth}
      \begin{algorithm2e}[H]
        \KwIn{$G = (V, E)$, $k$, $\varepsilon > 0$, const. $C$}
        \KwOut{$k'$-way partition $\Pi$ of $G$}
        \eIf(\tcp*[f]{recursion}){$\lvert V(G) \rvert > 2C$ and coarsening has not converged}{
          $G_c \coloneqq \FuncSty{coarsen}(G)$ \label{line:coarsen} \\
          $\Pi_c \coloneqq \FuncSty{partition}(G_c, k, \varepsilon, C)$ \label{line:recursion} \\
          $\Pi \coloneqq \FuncSty{project}(G, \Pi_c)$ \\
        }(\tcp*[f]{base case}){
          $\Pi \coloneqq \{V\}$ \\
        }

        $k' \coloneqq \left\{
        \begin{array}{ll}
          k, & \lvert V \rvert = n \\
          \textrm{ceil}_2(\lvert V \rvert / C), & \text{else}
        \end{array}
        \right.$ \label{line:ibp_start} \\
        $k' \coloneqq \max\{\min\{k, k'\}, 2\}$ \\
        \While(\tcp*[f]{obtain $k'$ blocks}){$\lvert \Pi \rvert < k'$}{
          $\Pi \coloneqq \FuncSty{bipartitionBlocks}(G, \Pi, k)$ \label{line:ibp_end} \\
        }
        $\Pi \coloneqq \FuncSty{refine}(G, \FuncSty{balance}(G, \Pi))$ \label{line:ref} \\
        \Return{$\Pi$} \\
        \caption{\normalsize $\FuncSty{partition}$}\label{alg:deep_mgp}
      \end{algorithm2e}
    \end{minipage}\hfill
    \begin{minipage}[t]{0.48\textwidth}
      \begin{algorithm2e}[H]
        \KwIn{$G$, $k'$-way partition $\Pi$, $k$, const. $R$}
        \KwOut{$2k'$-way partition $\Pi'$}
        $\Pi' \coloneqq \emptyset$ \\
        \ForEach{$V_i \in \Pi$}{
          $\varepsilon' \coloneqq \left( (1 + \varepsilon) \frac{c(V)}{\lvert \Pi \rvert c(V_i)} \right)^{1/\log_2(k / \lvert \Pi \rvert)} - 1$ \\
          $G_i \coloneqq \subgraph{V_i}$ \\
          \textit{// compute $R$ bipartitions of $V_i$ \\
            $\Pi_1, \dots, \Pi_R \coloneqq \FuncSty{bipartition}(G_i, \varepsilon', R)$ \\
            \textit{// select lowest (feasible) edge cut} \\
            $\Pi' \coloneqq \Pi' \cup \FuncSty{lowest}(G_i, \Pi_1, \dots, \Pi_R)$} \\
        }
        \caption{\normalsize $\FuncSty{bipartitionBlocks}$}\label{alg:bipartition_blocks}
      \end{algorithm2e}
    \end{minipage}
  \end{figure}

  \myparagraph{Deep Multilevel Graph Partitioning.}
  Deep MGP starts by coarsening the input graph $G_1 = (V_1, E_1)$, building a hierarchy of successively coarse
  representations $G_1, \dots, G_{\ell}$ of $G_1$.
  This is achieved by clustering each graph $G_i$ and contracting all clusters to build $G_{i + 1}$.
  Coarsening stops once the coarsest graph $G_{\ell}$ has at most $2C$ nodes, or the process converged.
  In \Cref{alg:deep_mgp}, this process is implemented in Lines~\ref{line:coarsen}--\ref{line:recursion}.

  From here, we start a sequence of the following operations: use recursive initial bipartition to subdivide the current
  graph into more blocks, possibly rebalance the partition and improve it using a $k$-way local improvement algorithm,
  project the partition onto the previous graph $G_{\ell - 1}$ and repeat the process on that graph.
  During these operations, we maintain the following key invariants:

  \textbf{(P)} A coarse graph $G_i$ is partitioned into
  $k_i \coloneqq \textrm{ceil}_2(\lvert V_i \rvert / C)$ blocks (bounded by $2$ and $k$).

  \textbf{(B)} A $k_i$-way partition of $G_i$ fulfills the balance constraint.

  An idealized coarsening algorithm produces a graph hierarchy where the number of nodes is halved between two levels
  and the coarsest graph has $2C$ nodes.
  In this case, it is sufficient to bipartition the coarsest graph $G_{\ell}$ once to fulfill invariant (P).
  To restore the invariant after uncoarsening (doubling the number of nodes in the current graph), each block of the
  current partition has to be bipartitioned once.

  In the more general case, where coarsening can shrink the number of nodes of a graph by a larger factor than $2$, we
  use recursive initial bipartitioning to maintain (P).
  More precisely, whenever the partition $\Pi_i$ of graph $G_i$ violates invariant (P), we recursively bipartition each
  block of $\Pi_i$ until we have $k_i$ blocks in total.
  In \Cref{alg:deep_mgp}, this process is implemented in Lines~\ref{line:ibp_start}--\ref{line:ibp_end}.
  Initial bipartitioning is implemented in \Cref{alg:bipartition_blocks}.

  Uncoarsening and initial bipartitioning can cause violations of invariant (B) (see below).
  If this is the case, we run a balancing algorithm afterwards.
  The resulting partition -- which satisfies invariants (P) and (B) -- is then improved using a $k$-way local
  improvement algorithm (\Cref{alg:deep_mgp}, Line~\ref{line:ref}).

  If $k > \textrm{ceil}_2(\lvert V_1 \rvert / C)$, the partition computed by the process described above has less than
  $k$ blocks.
  In this case, we perform an additional round of recursive initial bipartitioning to obtain $k$ blocks, and run
  balancing and $k$-way local improvement once more on the final partition.

  \myparagraph{Parallelization.}
  We parallelize the partitioning method described above using parallel coarsening, local improvement and balancing
  algorithms.
  On very coarse levels, we maintain the invariant that parallel tasks performed by $p$ PEs work on graphs with at least
  $pC$ nodes.
  This is achieved by running initial partitioning on more and more copies of the coarsened graph, as
  illustrated by \Cref{fig:part_scheme}.
  We diversify this search by using randomized coarsening, initial bipartitioning and local improvement algorithms.
  More precisely, we follow the algorithm described above until the coarsest graph $G_C$ has $pC$ nodes left.
  To uphold the invariant that tasks performed by $p$ PEs work on graphs with at least $pC$ nodes, we obtain two
  copies $G^r_C$ and $G^{\ell}_C$ of $G_C$, and split PEs into two groups (conceptually) with $p' = \frac{p}{2}$ PEs
  each.
  If $p' > 1$, we continue by coarsening $G^r_C$ with PEs of the first group and $G^{\ell}_C$
  with PEs of the second group, until each graph has $p'C$ nodes left.
  We proceed in this fashion recursively, until we have obtained $p$ graphs with $2C$ nodes each after $\log_2(p)$
  recursion levels.

  Each of these graphs is then bipartitioned using a single PE.
  Let $G^r_C$ and $G^{\ell}_C$ with respective bipartitions $\Pi^r_C$ and $\Pi^{\ell}_C$ be two such graphs that are
  copies of $G_C$ on the previous recursion level.
  We use the better bipartition of $\Pi^r_C$ and $\Pi^{\ell}_C$
  (\ie, if only one of these partitions is feasible, we use that one, otherwise the one with the lower edge cut)
  as partition $\Pi_C$ of $G_C$.
  We proceed on $G_C$ as before, \ie, bipartition each block of $\Pi_C$ if applicable, and apply the balancing and local
  improvement algorithm.
  This process is repeated for all $\log_2(p)$ recursion levels.

  \myparagraph{Handling General $k$.}
  The simplified description of deep MGP only considers the case where $k$ is a power of two.
  For the general case, we associate each block $B$ with a final block count $f_B$ -- the number of blocks $B$ is
  subdivided into in the final partition.
  Initially, $f_V = k$.
  To bipartition $B$ into two blocks $B_0$ and $B_1$, we set $f_{B_0} = \lfloor \frac{f_B}{2} \rfloor$ and
  $f_{B_1} = \lceil \frac{f_B}{2} \rceil$, and divide the weight of $B$ in a $f_{B_0}$ to $f_{B_1}$ ratio between $B_0$
  and $B_1$.
  Thus, once we have computed a $k' \coloneqq \textrm{floor}_2(k)$-way partition, there are $k - k'$ heavy blocks
  with $f_B = 2$ and $2k' - k$ light blocks with $f_B = 1$.
  During the next and final initial partitioning step, we obtain a $k$-way partition by only bipartition heavy
  blocks.

  \myparagraph{Maintaining the Balance Constraint.}
  Since MGP implementations usually employ coarsening algorithms that do not guarantee strictly uniform node weights,
  maintaining the balance constrain used in other partitioning systems,
  $L_k \coloneqq (1 + \varepsilon) \lceil \frac{c(V)}{k} \rceil$, becomes an NP-complete
  problem~\cite{garey1979computers} on coarse levels.
  To mitigate this problem, we use
  $L_{\max,k} \coloneqq \max\{ (1 + \varepsilon) \frac{c(V)}{k}, \frac{c(V)}{k} + \max_v c(v)\}$
  as balance constraint instead.
  This ensures that a feasible partition always exists, and that it can be found with simple greedy algorithms.
  Both claims are based on the fact that the average block weight of a partition is $\frac{c(V)}{k}$ and thus,
  there always exists a block $V_i$ with $c(V_i) \le \frac{c(V)}{k}$.
  In the multilevel setting, projecting a partition to a finer graph
  can violate the balance constraint due to the change in $\max_v c(v)$.
  However, the overload per block
  is bounded by $\max_v c(v)$, which implies that a balancing algorithm only needs to move a small
  number of nodes out of a block to restore the balance constraint.

  \myparagraph{Running Time.}
  Next, we analyze the running time of parallel (deep) MGP using
  highly idealized assumption.
  We do not claim that the results hold for our implementation but use this simplified analysis to give
  a quantitative expression to the qualitative reasoning that deep MGP is scalable if
  its components are scalable.
  The analysis also allows us to compare the
  asymptotic performance of different approaches to parallel MGP
  without having to discuss which particular implementations of the
  basic operations can or cannot avoid certain difficult cases.
  We assume:
  (1) $k$ is a power of two and we have unit node/edge weights,
  (2) $n>Cp\log p$,
  (3) coarsening a graph halves the number of nodes,
  (4) (un)coarsening or balancing a graph with $n$ nodes takes time $\Oh{n/p+\log n}$,
  (5) sequential bipartitioning takes linear time.
  We effectively ignore edges here.
  This implies the assumption that nodes have bounded degree and that the degrees
  remain bounded when the graph is shrunk.

  \begin{restatable}{theorem}{thmrunningtime}{}\label{thm:running_time}
    Under the assumptions made above, deep MGP requires
    time
    \[ \Oh{\frac{n}{p}\max\left(1,\log\frac{kC}{n}\right)+\log^2 n}\punkt \]
  \end{restatable}

  The proof can be found in \Cref{appendix:runtime_proof}.

  \section{Implementation}\label{sec:instantiation}

  In this section we describe the different components in our shared-memory parallel implementation of deep MGP called \Partitioner{KaMinPar}.
  Recall that the components are coarsening, bipartitioning on small graphs, and uncoarsening with $k$-way balancing and refinement.

  \subsection{Coarsening by Size-Constrained Label Propagation}\label{subsec:coarsening}
  We use size-constrained~label~propagation~\cite{LPAgraphPartitioning} to compute a clustering for contraction, where the weight of the
  heaviest cluster is restricted by a fixed upper bound $U$.
  We set $U \coloneqq \varepsilon \frac{c(V)}{k'}$,
  where $k' = \min\{k, \lvert V \rvert / C\}$ is the number of blocks we obtain on the finest level (before further bipartitioning if necessary) and $\varepsilon$ is the imbalance from the problem formulation.
  This choice implies that
  $\frac{c(V)}{k'} + \max_v c(v) \stackrel{\scriptscriptstyle \max_v c(v) \le U}{\scriptstyle \le} (1 + \varepsilon) \lceil \frac{c(V)}{k'} \rceil = L_{k'}$ on every level, and hence $L_{\max,k}$ simplifies to the traditional balance constraint $L_k$ on unweighted inputs.
  If the current number of nodes is $\leq \frac{k'}{2}C$, we adapt $k'$ to $\frac{k'}{2}$ and $U$ accordingly.
  We perform \numprint{5} rounds of label propagation, but terminate early if no nodes were moved.
  To further improve the running time, we only move a node, if one of its neighbors
  changed its cluster in the previous round.

  \myparagraph{Parallelization.}
  We parallelize the algorithm by iterating over all nodes in parallel.
  When moving a node to another cluster, we use atomic fetch-and-add operations to update the respective cluster
  weights.
  Note that we do not strictly enforce the weight limit.
  The limit could be violated if multiple PEs move a node to the same cluster at the same time.
  However, this is unproblematic in practice since the weight limit violations are usually small.

  \myparagraph{Iteration Order and Cache Locality.}
  Solution quality of label propagation is improved when nodes are visited in increasing degree
  order~\cite{LPAgraphPartitioning, DBLP:journals/tpds/AkhremtsevSS20}.
  Since this is not cache efficient and lacks diversification by randomization, we sort the nodes of the graph into exponentially spaced degree buckets, \ie, bucket $i$ contains all
  nodes with degree $2^i \le d < 2^{i + 1}$, and rearrange the graph such that nodes are sorted by their bucket number.
  For node traversal, we split buckets into small chunks and randomize node traversal on a inter-chunk
  and intra-chunk level.
  This is analogous to the randomization in \textsf{Metis}' matching algorithm~\cite{karypis1998fast}.

  \myparagraph{Two-hop Clustering.}
  We observed that size-constrained label propagation is unable to shrink some irregular graph instances sufficiently.
  We solve this by implementing a technique similar to the two-hop matching algorithm of
  \Partitioner{Metis}~\cite{DBLP:conf/sc/LaSallePSSDK15}.
  During label propagation, if node $u$ cannot be moved into any neighboring cluster due to the size constraint,
  we store the highest rated neighboring cluster as $u$'s \emph{favored cluster}.
  If the graph is shrunk by less than $50\%$ after termination, we merge singleton clusters that share the same favored cluster until the graph shrunk by $50\%$.

  \subsection{Initial Bipartitioning}\label{subsec:initial_bipartitioning}

  We perform multilevel bipartitioning to compute an initial bipartition of a subgraph $G_{V'}$
  (with $|V'| \approx 2C$).
  On this size, the used algorithms are sequential.
  For coarsening, we use label propagation and set the maximum cluster weight to the same value used in
  \Partitioner{KaHIP}~\cite{kaffpa}.
  The maximum block weight for bipartitioning is set to
  $L_2 \coloneqq (1 + \varepsilon') \lceil \frac{c(V')}{2} \rceil$, where $\varepsilon'$ is the adaptive imbalance as
  defined in \Cref{sec:preliminaries}.
  We coarsen until no further contractions are possible.
  For refinement, we use $2$-way FM~\cite{fiduccia1982lth}.
  We use a pool of simple algorithms to bipartition the coarsest graph, namely random bipartitioning, breadth-first
  searches and greedy graph growing~\cite{karypis1998fast}.
  We repeat each algorithm several times with different random seeds and select the
  bipartition with the lowest edge cut.
  Moreover, we use the adaptive algorithm selection technique of \Partitioner{Mt-KaHyPar}~\cite{MT-KAHYPAR-N-LEVEL}.

  \subsection{Uncoarsening}\label{subsec:balancing_refinement}
  After bipartitioning the blocks of a $\frac{k}{2}$-way partition, we use a $k$-way balancing algorithm to restore the balance constraint
  (if violated).
  Afterwards, we run a local improvement algorithm based on size-constrained label propagation to improve it.

  \myparagraph{Balancing.}
  In contrast to \Cref{sec:deep_mgp}, our implementation prevents balance constraint violations by changes in
  $\max_v c(v)$ due to our choice of the maximum cluster weight during coarsening.
  However, balance violations can occur during initial bipartitioning, in particular due to our multilevel bipartitioning approach.

  For each overloaded block $B$, we store just enough nodes of $B$ in a priority queue $P_B$ ordered by \emph{relative gains},
  to remove the excess weight $o(B) \coloneqq c(B) - L_{\max,k}$.
  The \emph{relative gain} of a node $v$ is $d \cdot c(v)$ if $d \geq 0$ and $d / c(v)$ if $d < 0$,
  where $d$ is the largest reduction in edgecut when moving $v$ to a block that would not become overloaded.
  We initialize the priority queues by iterating over the nodes in $G$.
  If a node is in an overloaded block and $c(P_B) < o(B)$, we insert it.
  Otherwise, we only insert it if its relative gain is higher than the lowest relative gain of any element in $P_B$ and
  remove its lowest element if $c(P_B) > o(B) + \max_v c(V)$ after the insertion.

  Once the priority queues are initialized, we empty each overloaded block $B$ individually by repeatedly removing the node $v$
  with the largest relative gain from $P_B$.
  If its relative gain changed since insertion, or its designated target block can no longer take $v$ without becoming overloaded,
  we re-insert $v$ (if $v$ is still a border node).
  Otherwise, we move $v$ to its target block or a random block that can take $v$ without
  becoming overloaded.
  Subsequently, we try to insert all neighbors from its former block.
  To reduce the running time, we only try to insert each node once.

  We parallelize the algorithm as follows.
  During initialization, we iterate over all nodes in parallel and maintain one thread-local priority queue for each
  overloaded block.
  Afterwards, we iterate over all blocks in parallel, merge the respective thread-local priority queues and perform
  node movements as described above.

  \myparagraph{Local Improvement.}
  We use the same parallelization of size-constrained label propagation as described in \Cref{subsec:coarsening}, but strictly enforce the maximum cluster weight (set to the maximum block weight) using an atomic compare-and-swap instruction.
  We run at most \numprint{5} rounds of size-constrained label propagation (same value as used in \Partitioner{Mt-KaHyPar}~\cite{MT-KAHYPAR-FAST}), but terminate early if no node was moved
  during a round.

  \section{Experimental Evaluation}\label{sec:eval}

  We implemented the proposed algorithm \Partitioner{KaMinPar}
  in C++ and compiled it using g++-10.2 with flags \texttt{-O3 -march=native}.
  We use Intel's TBB~\cite{TBB} as parallelization library.

  \myparagraph{Setup.}
  We perform our experiments on two different machines.
  Machine \textsf{A} is equipped with an AMD EPYC 7702 64-Core processor clocked at 2 GHz and 1 TB main memory.
  This machine is only used for our scalability experiment.
  All other experiments are run on Machine \textsf{B}, which is a node of a cluster equipped with Intel Xeon Gold 6230
  processors (2 sockets with 20 cores each) clocked at 2.1 GHz and 96 GB or 192 GB main memory.

  We compare our algorithm with
  \Partitioner{Mt-Metis 0.7.2}~\cite{lasalle2016parallel},
  \Partitioner{Mt-KaHiP 1.0}~\cite{DBLP:journals/tpds/AkhremtsevSS20},
  \Partitioner{PuLP 0.11}~\cite{pulp},
  \Partitioner{Metis 5.1.0}~\cite{DBLP:conf/sc/LaSallePSSDK15} and the \Partitioner{fsocial} preset of
  \Partitioner{KaHiP 3.10}~\cite{kaffpa}.
  We chose this preset because it is one of the fastest configurations that computes good quality.
  While other presets of \Partitioner{KaHiP} achieve better partition quality, they are also much slower.
  We do not include \Partitioner{ParMetis}~\cite{karypis1996parallel} and \Partitioner{Pt-Scotch}~\cite{ptscotch} in our comparison since they are
  slower than \Partitioner{Mt-Metis} and produce partitions with comparable solution quality~\cite{DBLP:conf/ipps/LasalleK13}.
  Moreover, we exclude \Partitioner{ParHiP}~\cite{DBLP:journals/tpds/MeyerhenkeSS17} since it is outperformed by \Partitioner{Mt-KaHiP}~\cite{Akhremtsev2019_1000098895}.
  In the following, we add a suffix to the name of each parallel partitioner to indicate the number of threads used, e.g.,~\Partitioner{KaMinPar 64} for \numprint{64} threads.

  \myparagraph{Instances.}
  We evaluate our algorithm on a benchmark set composed of \numprint{197} graphs (referred to as set A), including
  \numprint{129} graphs from the 10th~DIMACS~Implementation~Challenge~\cite{dimacschallengegraphpartandcluster},
  \numprint{25} randomly generated graphs~\cite{funke2017communication,wsvr}, \numprint{25} large social networks~\cite{webgraphWS,snap}, and
  \numprint{18} graphs from various application domains~\cite{FloridaSPM,wovsyd2007,DAC2012}.
  Scalability and experiments with larger values of $k$ are performed on a subset of set \textsf{A} that contains \numprint{21} graphs
  (referred to as set \textsf{B}). This benchmark set includes the \numprint{18} largest graphs
  (by number of nodes)\footnote{excluding \texttt{er-fact1.5-scale26}, since \Partitioner{Mt-KaHiP} and \Partitioner{Mt-Metis} are unable to compute
  a partition on this graph even for small $k$, and \texttt{kmer\_V2a} to avoid over-representation of k-mer graphs} and
  \numprint{3} randomly chosen small graphs of set \textsf{A} such that a partition
  with $2^{20}$ blocks only contains a few nodes per block.
  Basic properties of benchmark instances are shown in \Cref{sec:apx:benchmark_stats}, \Cref{fig:benchmark_stats}.

  \myparagraph{Methodology.}
  We consider a combination of a graph and number of blocks $k$ as an \emph{instance}. For each instance,
  we usually perform several runs with different random seeds and aggregate running times and edge cuts using
  the arithmetic mean over all seeds. To further aggregate over multiple instances, we use the harmonic mean for
  relative speedups, and the geometric mean for absolute running times and edge cuts. Runs with imbalanced
  partitions are not excluded from aggregated running times and for instances that exceeded
  the time limit, we use the time limit in the aggregates. We consider an instance as infeasible, if
  all runs failed or computed an imbalanced partition and mark them with \ding{55} in the plots.

  To compare the solution quality of different algorithms, we use \emph{performance profiles}~\cite{DBLP:journals/mp/DolanM02}.
  Let $\mathcal{A}$ be the set of all algorithms we want to compare, $\mathcal{I}$ the set of instances, and $q_{A}(I)$ the quality of algorithm
  $A \in \mathcal{A}$ on instance $I \in \mathcal{I}$.
  For each algorithm $A$, we plot the fraction of instances $\frac{|\mathcal{I}_{A}(\tau)|}{|\mathcal{I}|}$ ($y$-axis) where
  $\mathcal{I}_{A}(\tau) \coloneqq \{ I \in \mathcal{I}~|~q_A(I) \leq \tau \cdot \min_{A' \in \mathcal{A}}q_{A'}(I) \}$ and $\tau$ is on the $x$-axis.
  Achieving higher fractions at lower $\tau$-values is considered better.
  For $\tau = 1$, the $y$-value indicates the percentage of instances for which an algorithm performs best.
  Since performance profiles relate the quality of an algorithm to the best solution,
  the ranking induced by $\tau = 1$ does not permit full ranking of all algorithms,
  if more than two algorithms are included.

  \myparagraph{Running Time and Solution Quality for Small $k$.}
  In Figure~\ref{fig:defaultk_results} and \Cref{tbl:defaultk_results} in Appendix~\ref{sec:apx:defaultk_results},
  we compare the quality and running time of \Partitioner{KaMinPar} with different partitioners for $k \in \{2,4,8,16,32,64\}$,
  $\varepsilon = 0.03$ and $5$ repetitions per instance on set \textsf{A} and machine \textsf{B}.
  These are commonly used values to evaluate graph partitioning systems.
  Further, we execute each parallel partitioner using $10$ threads to simulate the performance of them on commodity machines.

  \Partitioner{KaMinPar 10} is the fastest algorithm on average
  and also an order of magnitude faster than the sequential partitioners \Partitioner{Metis}
  and \Partitioner{KaHIP-fsocial} on large graphs ($m \ge 10^8$), while producing partitions with
  comparable solution quality (see \Cref{fig:smallk_additional_edge_cut} (left) in \Cref{sec:apx:defaultk_results}). \Partitioner{KaMinPar 10} (\numprint{0.39} s geometric mean running time) is moderately faster
  than \Partitioner{Mt-Metis 10} (\numprint{0.48} s) and more than a factor of \numprint{2} resp.~\numprint{3} faster than \Partitioner{PuLP 10} (\numprint{1.11} s)
  and \Partitioner{Mt-KaHIP 10} (\numprint{1.33} s). The differences in running time, as shown in Figure~\ref{fig:defaultk_results} (right), are more
  pronounced on larger instances, e.g., \Partitioner{KaMinPar 10} (\numprint{9.36} s) is more than factor of \numprint{3} resp.~\numprint{5} faster than \Partitioner{Mt-Metis 10} (\numprint{30.36} s)
  resp.~\Partitioner{Mt-KaHIP 10} (\numprint{55.76} s) on instances with more than $10^8$ edges. Figure~\ref{fig:defaultk_results} (left) shows that
  \Partitioner{Mt-KaHIP 10} computes on most of the instances the partition with lowest edge cut ($\approx$\,\numprint{60}\% of the instances), while
  the partitions produced by \Partitioner{PuLP 10} are more than a factor of \numprint{2} worse than the best achieved edge cuts on more than \numprint{55}\% of the instances.
  These results are expected, since \Partitioner{Mt-KaHIP} is the only partitioner that implements a parallel direct $k$-way FM algorithm
  and \Partitioner{PuLP} is the only non-multilevel system in our evaluation. We perform the same experiment using $64$ cores of machine \textsf{A}
  on the larger instances of set \textsf{B} with similar results (see Appendix~\ref{sec:apx:defaultk_results}).
  Thus, we can conclude that \Partitioner{KaMinPar} offers a compelling trade-off between running time and quality compared to
  established shared-memory and sequential GP systems.

  \begin{figure}
    \begin{subfigure}[t]{0.5\textwidth}
      \vspace*{0em}
      
\ifextfig
  \input{tr_plots/smallk_shm_edge_cut}
\else
  \includegraphics{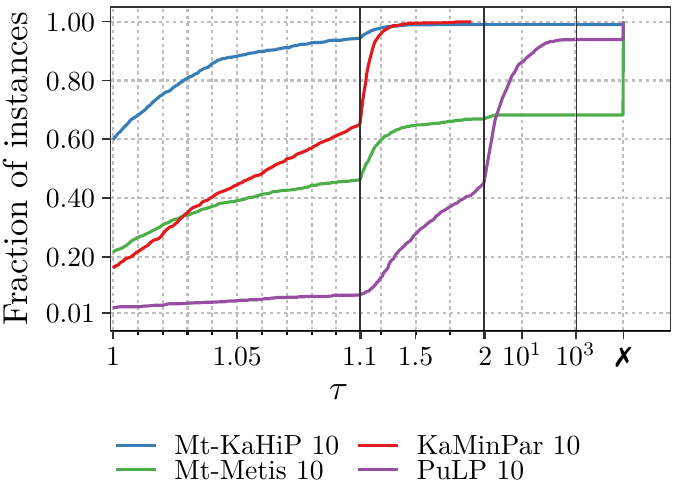}
\fi

    \end{subfigure}
    \begin{subfigure}[t]{0.5\textwidth}
      \vspace*{-0.03em}
      
\ifextfig
  \input{tr_plots/smallk_running_time}
\else
  \includegraphics{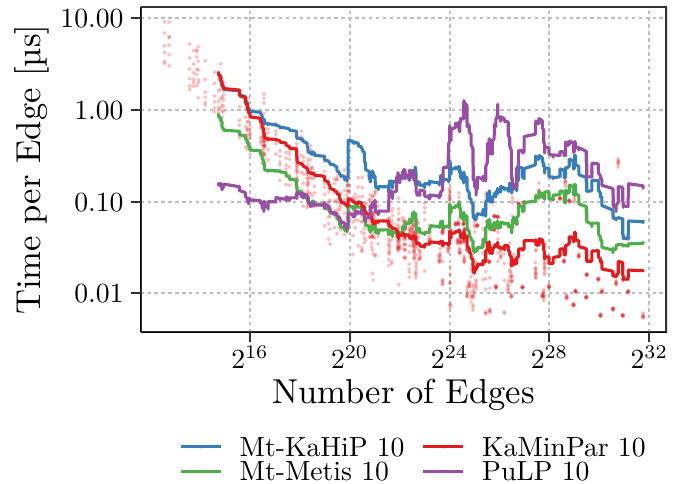}
\fi

    \end{subfigure}
    \ifextfig
    \vspace*{-1.5em}
    \fi
    \caption{ Performance profile and running time plot (shows time per edge with a right-aligned rolling geometric mean over \numprint{50} instances)
      comparing the performance of \Partitioner{KaMinPar} with different partitioners for smaller values of $k$ on set A.}
    \label{fig:defaultk_results}
    \ifextfig
    \vspace*{-0.25cm}
    \else
    \vspace*{-0.3cm}
    \fi
  \end{figure}

  \begin{table}[t]
    \centering
    \caption{Results of our experiment for large values of $k$ with different
      parallel partitioners on set \textsf{B}. The last two columns show the geometric mean running time
      and edge cuts relative to \Partitioner{KaMinPar} of all instances that do not crash (timeout instances
      are additionally excluded in edge cut comparison).}
    \label{tbl:largek_results}
    \vspace{-0.25cm}
    \begin{tabular}{l|rrrr|rr}
      Algorithm & \multicolumn{1}{c}{\# timeout} & \multicolumn{1}{c}{\# crash} & \multicolumn{1}{c}{\# imbalanced} & \multicolumn{1}{c|}{\# feasible} & \multicolumn{1}{c}{rel. time} & \multicolumn{1}{c}{rel. cut} \\
      \midrule
      \Partitioner{KaMinPar~10} & \textbf{\numprint{0}} & \textbf{\numprint{0}} & \textbf{\numprint{0}} & \textbf{\numprint{84}} & \numprint{1.00} & \numprint{1.00} \\
      \midrule
      \Partitioner{Mt-Metis-K~10} & \numprint{19} & \numprint{10} & \numprint{51} & \numprint{4} & \numprint{11.91} & \textbf{\numprint{0.99}} \\
      \Partitioner{Mt-Metis-RB~10} & \textbf{\numprint{0}} & \numprint{25} & \numprint{55} & \numprint{4} & \numprint{5.61} & \numprint{1.03} \\
      \Partitioner{Mt-KaHiP~10} & \numprint{31} & \numprint{7} & \numprint{11} & \numprint{35} & \numprint{38.64} & \numprint{1.00} \\
      \Partitioner{PuLP~10} & \numprint{76} & \numprint{0} & \textbf{\numprint{0}} & \numprint{8} & \numprint{73.52} & \numprint{1.25} \\
    \end{tabular}
    \vspace{-0.5cm}
  \end{table}

  \myparagraph{Running Time and Solution Quality for Large $k$.}
  In \Cref{tbl:largek_results},
  we present the results of our experiment with different parallel partitioners (each using $10$ threads) for larger values of $k \in \{2^{11}, 2^{14}, 2^{17}, 2^{20}\}$,
  $\varepsilon = 0.03$ and \numprint{3} repetitions per instance with a time limit of one hour on set \textsf{B} and machine \textsf{B} (\numprint{192} Gb main memory).
  Note that the time limit is \numprint{10} times larger than the longest running time of \Partitioner{KaMinPar} for an instance.
  We additionally included the recursive bipartitioning version of \Partitioner{Mt-Metis} (referred to as \Partitioner{Mt-Metis-RB})
  to evaluate the performance of recursive bipartitioning on large $k$.
  In the following, we consider a partition produced by an algorithm as feasible, if the algorithm terminates in the  given time limit and
  satisfies the balance constraint $L_k \coloneqq (1 + \varepsilon) \lceil \frac{c(V)}{k} \rceil$.

  Out of the \numprint{84} evaluated instances (\numprint{21} graphs times \numprint{4} values of $k$), \Partitioner{Mt-Metis-RB 10}, \Partitioner{Mt-Metis-K 10}, \Partitioner{PuLP 10} and
  \Partitioner{Mt-KaHIP 10} were only able to produce on \numprint{4}, \numprint{4}, \numprint{8} resp.~\numprint{35} instances a feasible solution.
  \Partitioner{Mt-Metis-RB 10} and \Partitioner{Mt-Metis-K 10} primarily failed to produce solutions that satisfy the
  balance constraint (\numprint{55} resp.~\numprint{51} instances).
  \Cref{fig:allk_imbalance} (right) in Appendix~\ref{sec:apx:infeasible} shows that \Partitioner{Mt-Metis-K 10}
  generally produces larger balance violations (median resp.~maximum imbalance is \numprint{1.14} resp.~\numprint{15.56}) than \Partitioner{Mt-Metis-RB 10}
  (median \numprint{1.05} and maximum \numprint{1.15}).
  \Partitioner{Mt-KaHIP 10} and \Partitioner{PuLP 10} were mostly unable to compute a partition in the given time limit
  (\numprint{31} resp.~\numprint{76} instances).
  \Partitioner{KaMinPar 10} produced a feasible solution on all instances.
  The fastest competitor is \Partitioner{Mt-Metis-RB 10}, which is more than \numprint{5} times slower than
  \Partitioner{KaMinPar 10} on average.
  All other partitioners are an order of magnitude slower than \Partitioner{KaMinPar 10}.
  If we include all infeasible partitions and individually compare the partitioners on those instances
  in terms of solution quality, we can see that all produce partitions with comparable edge cuts
  (except for \Partitioner{PuLP 10}).
  However, a fair comparison in terms of solution quality is difficult due to the large number of infeasible solutions.
  We conclude that \Partitioner{KaMinPar} is currently the only partitioner considered in our evaluation that can
  reliably compute feasible partitions for larger values of $k$ in a reasonable amount of time.

\begin{figure}
    \centering
    
\ifextfig
  \input{tr_plots/largek_scaling}
\else
  \includegraphics{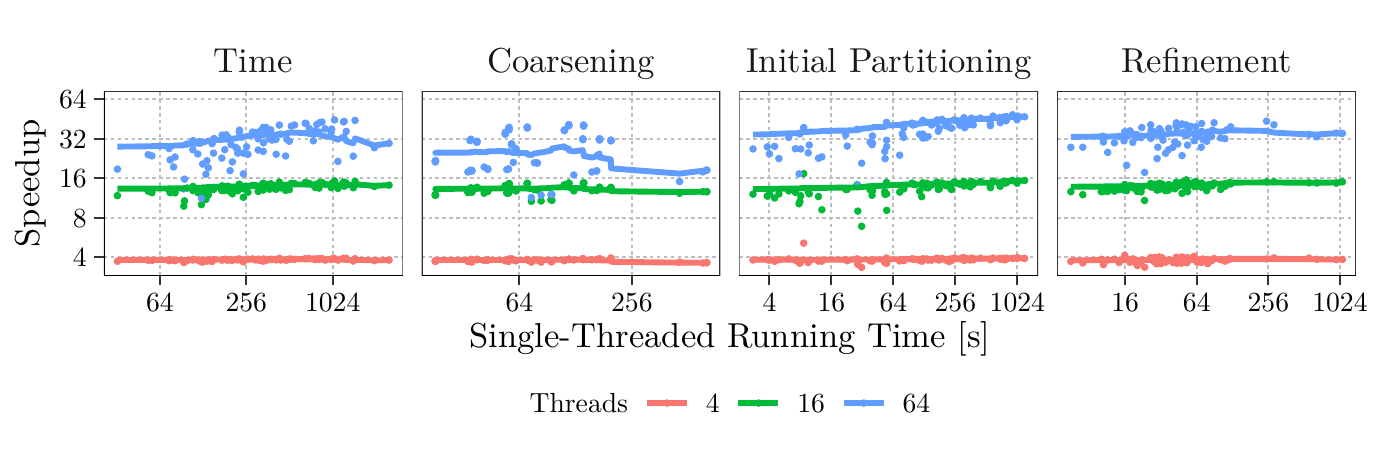}
\fi

    \vspace*{-2em}
    \caption{Self-relative speedups for the different components of \Partitioner{KaMinPar} on set \textsf{B}.}
    \vspace*{-1em}
    \label{fig:scaling}
  \end{figure}

  \myparagraph{Scalability of KaMinPar.}
  In Figure~\ref{fig:scaling}, we show the scalability of \Partitioner{KaMinPar} for
  $k \in \{2^{11}, 2^{14}, 2^{17}, 2^{20}\}$, $\varepsilon = 0.03$ and three repetitions
  per instance on set \textsf{B} using $p \in \{1,4,16,64\}$ cores of machine \textsf{A}.
  In the plot, we represent the speedup of each instance as a point and the cumulative harmonic mean
  speedup over all instances with a single-threaded running time $\ge x$ seconds with a line.
  Note that initial partitioning includes all calls to our initial bipartitioning algorithms on graphs
  with more than $2pC$ nodes.

  The overall harmonic mean speedup of \Partitioner{KaMinPar} is \numprint{3.8} for $p = 4$, \numprint{13.3} for
  $p = 16$ and \numprint{27.9} for $p = 64$.
  The harmonic mean speedups of coarsening, initial partitioning and refinement are \numprint{25.0}, \numprint{34.5} and
  \numprint{33.1} for $p = 64$.
  We note that our refinement component achieves slightly better speedups than our coarsening component, although both
  are based on the size-constrained label propagation algorithm.
  This effect is most pronounced on instances with larger node degrees.
  During coarsening, each thread aggregates ratings to neighboring clusters in a local hash map (with only
  $2^{15}$ entries) for nodes with small degree and in a local vector of size $n$ for high degree nodes
  ($\ge 2^{15}/3$).
  During refinement, each thread uses a local vector of size $k$ for this.
  Instances with larger node degrees more often uses the local vector of size $n$ to aggregate ratings during
  coarsening, which can limit scalability due to cache effects.
  Note that \Partitioner{KaMinPar} performs no expensive arithmetic operations.
  Hence, perfect speedups are not possible due to limited memory bandwidth.

  \section{Conclusion and Future Work}\label{sec:conclusion}
  We presented a new graph partitioning scheme that successfully combines the merits of classical direct $k$-way
  partitioning and recursive bipartitioning.
  Similar to direct $k$-way partitioning, deep MGP coarsens and uncoarsens the graph only once, and allows the use of
  $k$-way local improvement algorithms.
  Yet, it does not suffer scalability problems if $k$ is large and has a better asymptotic running time than
  recursive bipartitioning.
  Our experimental evaluation shows that our shared-memory parallel implementation of deep MGP runs efficiently on up to
  \numprint{64} PEs, while achieving comparable results to established graph partitioners if $k$ is small.
  Furthermore, our evaluation showed that \Partitioner{KaMinPar} is an order of magnitude faster than other graph
  partitioners based on direct $k$-way partitioning if $k$ is large, while consistently producing balanced solutions.
  In the future, we would like to explore graph partitioning for very large values of $k$, e.g., $k \in \Theta(n)$.

  \clearpage
  \appendix
  \bibliography{phdthesiscs,bibliography,diss,p50-schlag}
  \newpage

  \begin{appendix}
    \section{Parallel Multilevel Graph Partitioning}
    \label{appendix:parallel_graph_partitioning}

    Graph partitioners that are able to partition large real-world graphs can be categorized into three types of
    systems: streaming~\cite{DBLP:conf/kdd/StantonK12,DBLP:journals/corr/abs-2102-09384,
    DBLP:conf/wsdm/TsourakakisGRV14,DBLP:journals/pvldb/AbbasKCV18,DBLP:conf/icde/MartellaLLS17,
    DBLP:conf/kdd/NishimuraU13}, non-multilevel~\cite{DBLP:journals/siamsc/SlotaMR16, DBLP:conf/ipps/SlotaRDM17}
    and parallel multilevel algorithms \cite{pulp,DBLP:conf/ipps/SlotaRDM17,ptscotch,karypis1996parallel,DBLP:journals/tpds/AkhremtsevSS20,DBLP:journals/tpds/MeyerhenkeSS17}.
    Streaming graph partitioners operate in a model in which the nodes $v$ and their neighborhood arrive one at a time and the
    algorithm directly has to assign $v$ to a block.
    Non-multilevel systems such as PuLP~\cite{pulp,DBLP:conf/ipps/SlotaRDM17} directly partition the input graph into $k$
    blocks and use local search heuristics to further improve solution quality.
    Streaming and non-multilevel systems significantly sacrifice solution quality to achieve higher
    speed~\cite{DBLP:conf/wsdm/TsourakakisGRV14,DBLP:conf/icde/MartellaLLS17,DBLP:journals/siamsc/SlotaMR16,DBLP:journals/corr/abs-2102-09384}.

    The matching~\cite{ptscotch, Zoltan, lasalle2013multi, karypis1996parallel, Walshaw07} and
    clustering algorithms~\cite{DBLP:journals/tpds/AkhremtsevSS20, PARALLEL-PATOH,MT-KAHYPAR-FAST}
    used by different sequential partitioners in the coarsening phase are
    well-suited for parallelization, sometimes with only minor quality losses~\cite{PARALLEL-PATOH,BOSSS13}.
    The coarsening phase proceeds until a fixed number of nodes remains, usually $kC$, where $C$ is
    a tuning parameter.

    In the initial partitioning phase, parallel partitioners either
    call sequential multilevel algorithms with different random
    seeds~\cite{DBLP:journals/tpds/AkhremtsevSS20, ptscotch, Zoltan, kappa} or use
    parallel recursive bipartitioning~\cite{MT-KAHYPAR-FAST, lasalle2013multi, karypis1996parallel, lasalle2016parallel}. In the former case, the graph
    is copied to each PE and the best partition obtained from all independent runs is projected back
    to the coarsest graph.
    In the latter case, parallelism is achieved by either splitting the thread pool into two evenly-sized groups and assigning each to one of the two recursive
    calls~\cite{karypis1996parallel,lasalle2013multi} or dynamically assign the threads to the recursive calls with a task scheduler~\cite{MT-KAHYPAR-FAST}.
    To obtain an initial bipartition of the coarsest graph, often portfolio approaches
    composed of different bipartitioning algorithms are used~\cite{kaffpa, KaHyPar-R, MT-KAHYPAR-FAST, catalyurek1999hypergraph}.

    Most parallel partitioners use the label propagation heuristic to improve the solution quality during
    the refinement phase~\cite{DBLP:journals/tpds/AkhremtsevSS20, lasalle2013multi, MT-KAHYPAR-FAST, DBLP:journals/tpds/MeyerhenkeSS17, Walshaw07, Zoltan}.
    More advanced techniques are based on parallel
    variants of the FM local search~\cite{fiduccia1982lth} that are
    widely used in sequential partitioners.
    \Partitioner{PT-Scotch}~\cite{ptscotch}, \Partitioner{KaPPa}~\cite{kappa} and
    \Partitioner{Jostle}~\cite{Walshaw07} use sequential $2$-way FM refinement
    on two adjacent blocks of the
    partition. \Partitioner{Mt-KaHIP}~\cite{DBLP:journals/tpds/AkhremtsevSS20}
    and \Partitioner{Mt-KaHyPar}~\cite{MT-KAHYPAR-FAST} implement a parallel
    $k$-way FM algorithm based on the \emph{localized multi-try} FM of the
    sequential graph partitioner \Partitioner{KaHIP}~\cite{kaffpa}.  Mt-Metis~\cite{lasalle2016parallel} uses greedy
    refinement (FM with only positive gain moves), and hill-scanning,
    a simplification of localized FM where small groups of vertices, whose individual gains are negative,
    are moved if their combined gain is positive.

    In the parallel setting, nodes can change their block concurrently which requires synchronization to ensure that the
    balance constraint is not violated~\cite{lasalle2013multi}.
    Existing systems either explicitly communicate their local changes
    and reject moves that would violate the balance
    constraint~\cite{lasalle2013multi, karypis1996parallel, Zoltan}
    or use atomic \emph{compare-and-swap} instructions to maintain block
    weights~\cite{MT-KAHYPAR-FAST, DBLP:journals/tpds/AkhremtsevSS20}.

    \section{Proof of \Cref{thm:running_time}}\label{appendix:runtime_proof}
    \begin{proof}
      By (3), MGP goes through $\log(n/C)$ levels so that the overhead
      terms ``$\log n$'' in (4) sum to $\Oh{\log^2n}$ -- we ignore these overhead from now on. While $> pC$
      nodes are left, the graph shrinks geometrically with the levels so
      that the total remaining cost for (un)coarsening and balancing
      from (4) is linear -- $\Oh{n/p}$.

      When $\leq pC$ nodes are left, replication and selection of the
      best partition keeps the number of nodes at each level at
      $\Th{pC}$. There are $\log p$ such levels incurring total cost
      $\Oh{C\log p}$ for (un)coarsening and balancing. By (2) this cost
      is bounded by $\Oh{n/p}$.

      For the cost of bipartitioning we consider three cases:\\
      {\bf Case (a) $k\leq p$:} Each PE performs $\log k$ bipartitions with total cost $C\log k$.
      By (2) this is bounded by $\Oh{n/p}$.\\
      {\bf Case (b) $p < k \leq n/C$:} Each PE performs $\log p+k/p$ bipartitions with total cost $C(k/p+\log p)$.
      Once more, by (2) this is bounded by $\Oh{n/p}$.\\
      {\bf Case (c) $k>n/C$:} In this case, deep MGP first performs an $n/C$-way partitioning into blocks of size
      about $C$. By the above analysis, this takes time $\Oh{n/p+\log^2 n}$.
      Then the remaining blocks are partitioned into $k/(n/C)=kC/n$ blocks using recursive bipartitioning in time $\Oh{C\log(kC/n)}$.
      Summing over all blocks assigned to a PE we get additional cost $n\log(kC/n)/p$.
    \end{proof}

    \section{Benchmark Set Statistics}\label{sec:apx:benchmark_stats}

    \begin{figure}[H]
      \centering
      
\ifextfig
  \input{tr_plots/benchmark_stats}
\else
  \includegraphics{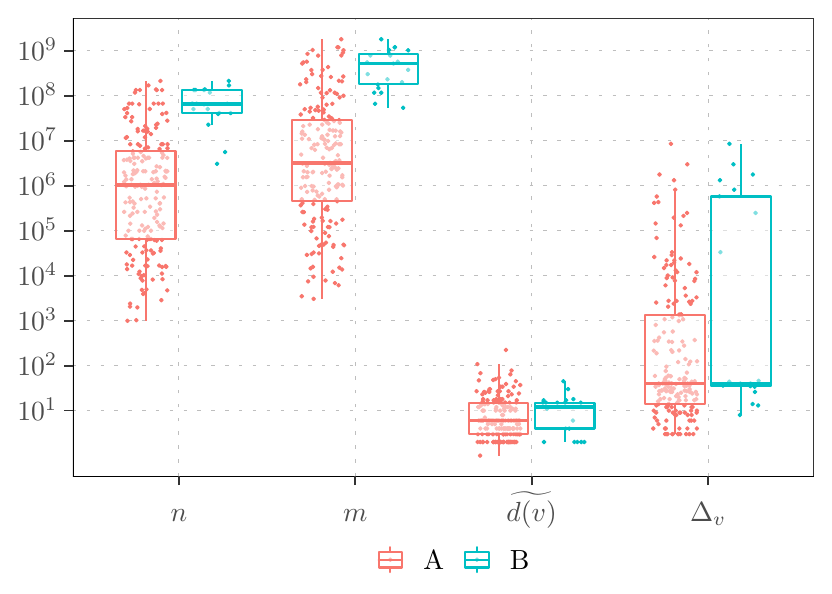}
\fi

      \caption{Basic properties of benchmark sets \textsf{A} and \textsf{B}: number of nodes $n$, number of edges $m$, median degree
        $\meddeg$ and maximum degree $\maxdeg{v}$.}
      \label{fig:benchmark_stats}
    \end{figure}

    \section{Running Time and Solution Quality with Small $k$}\label{sec:apx:defaultk_results}

\begin{figure}
      \centering
      \begin{subfigure}[t]{0.49\textwidth}
        \vspace*{0em}
        
\ifextfig
  \input{tr_plots/smallk_seq_edge_cut}
\else
  \includegraphics{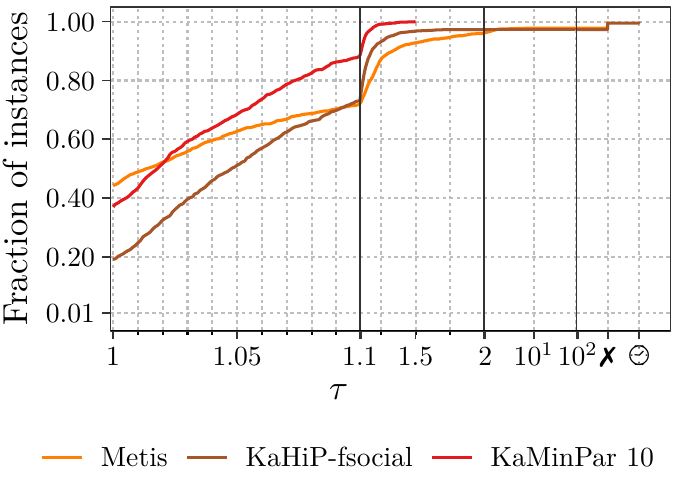}
\fi

      \end{subfigure}
      \begin{subfigure}[t]{0.49\textwidth}
        \vspace*{0em}
        
\ifextfig
  \input{tr_plots/smallk_t64_edge_cut}
\else
  \includegraphics{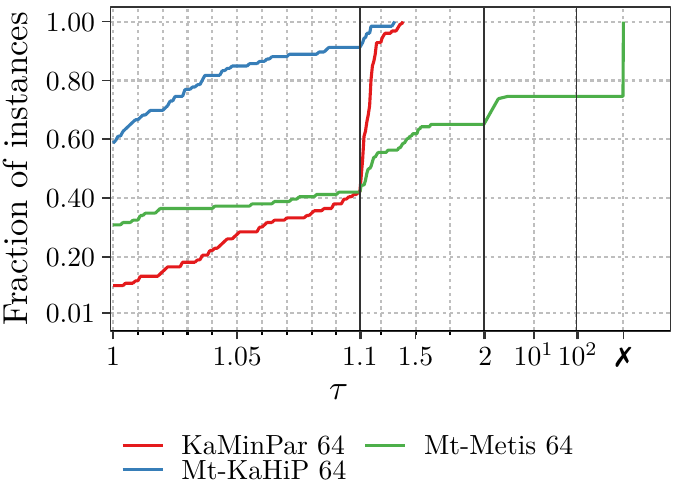}
\fi

      \end{subfigure}
      \caption{Left: performance profile of \Partitioner{KaMinPar}, \Partitioner{Metis} and \Partitioner{KaHiP-fsocial} on
      benchmark set \textsf{A} with $k \in \{2, 4, 8, 16, 32, 64\}$ and $\varepsilon = 0.03$.
      Right: performance profile of \Partitioner{KaMinPar}, \Partitioner{Mt-KaHiP} and \Partitioner{Mt-Metis} on
      \numprint{64} cores of machine \textsf{A}, reduced benchmark set \textsf{B} with $k \in \{2, 4, 8, 16, 32, 64\}$
      and $\varepsilon = 0.03$.
      Note that the change in relative solution quality is due to the reduced benchmark set.}
      \label{fig:smallk_additional_edge_cut}
    \end{figure}

    We also ran \Partitioner{KaMinPar}, \Partitioner{Mt-Metis} and
    \Partitioner{Mt-KaHiP} on all \numprint{64} cores of machine \textsf{A}.
    Here, we partitioned each graph of the reduced benchmark set \textsf{B} into $k \in \{2, 4, 8, 16, 32, 64\}$
    blocks allowing a maximum imbalance of $\varepsilon = 0.03$.
    In this setup, \Partitioner{KaMinPar 64} is \numprint{5} resp.~\numprint{4.4} times faster than \Partitioner{Mt-KaHiP 64} resp.~\Partitioner{Mt-Metis 64},
    whereas on the same instances on \numprint{10} cores of machine \textsf{B}, it is \numprint{5.7} resp.~\numprint{3.1} times faster.
    Thus, we can see that the running time of \Partitioner{KaMinPar} scales slightly worse to \numprint{64} cores than
    \Partitioner{Mt-KaHiP}, but slightly better than \Partitioner{Mt-Metis} for smaller values of $k$.
    In \Cref{fig:smallk_additional_edge_cut} (right), we compare the solution quality of each algorithm on the
    reduced benchmark set \textsf{B}. We can see that the differences of the partitioners in terms of solution quality
    are more pronounced on this set than on set \textsf{A} (compare with \Cref{fig:defaultk_results}).
    However, this is due to the benchmark set, since each partitioner produced partitions with comparable quality if
    we compare them individually with $10$ and $64$ threads.

    \begin{table}[H]
      \centering
      \caption{Geometric mean running time and solution quality for different algorithms on benchmark set \textsf{A} and
        $k \in \{2, 4, 8, 16, 32, 64\}$.
        Running time only includes instances for which all algorithms produced a result.
        The number of included instances is shown in the last row.
        Solution quality is relative to \Partitioner{KaMinPar} (lower is better) and only includes instances for which the
        respective algorithm computed a balanced partition.
        Thus, solution quality cannot be compared between different competitors.}
      \label{tbl:defaultk_results}
      \begin{tabular}{l|rrr|rr}
        Algorithm & \multicolumn{1}{c}{$T$} & \multicolumn{1}{c}{$T[m \ge 10^6]$} & \multicolumn{1}{c|}{$T[m \ge 10^8]$} & \multicolumn{1}{c}{rel. cut} & \multicolumn{1}{c}{\# infeasible} \\
        \midrule
        \Partitioner{KaMinPar~10} & \textbf{\numprint{0.39} s} & \textbf{\numprint{0.85} s} & \textbf{\numprint{9.36} s} & \numprint{1.00} & \textbf{\numprint{0}} \\
        \midrule
        \Partitioner{Mt-Metis~10} & \numprint{0.48} s & \numprint{1.49} s & \numprint{30.36} s & \numprint{1.00} & \numprint{349} \\
        \Partitioner{Mt-KaHiP~10} & \numprint{1.33} s & \numprint{3.84} s & \numprint{55.76} s & \textbf{\numprint{0.94}} & \numprint{6} \\
        \Partitioner{PuLP~10} & \numprint{1.11} s & \numprint{5.70} s & \numprint{95.93} s & \numprint{2.39} & \numprint{72} \\
        \midrule
        \Partitioner{Metis} & \numprint{1.00} s & \numprint{4.15} s & \numprint{97.44} s & \numprint{1.05} & \numprint{2} \\
        \Partitioner{KaHiP-fsocial} & \numprint{2.93} s & \numprint{11.05} s & \numprint{200.67} s & \numprint{1.03} & \numprint{8} \\
        \midrule
        \# instances & \numprint{1150} & \numprint{832} & \numprint{196} & \\
      \end{tabular}
    \end{table}

    \section{Infeasible Solutions}\label{sec:apx:infeasible}

    \begin{figure}[H]
      \begin{subfigure}[t]{0.5\textwidth}
        \vspace*{-0.05em}
        \hspace*{-2.9em}
        
\ifextfig
  \input{tr_plots/smallk_shm_imbalance}
\else
  \includegraphics{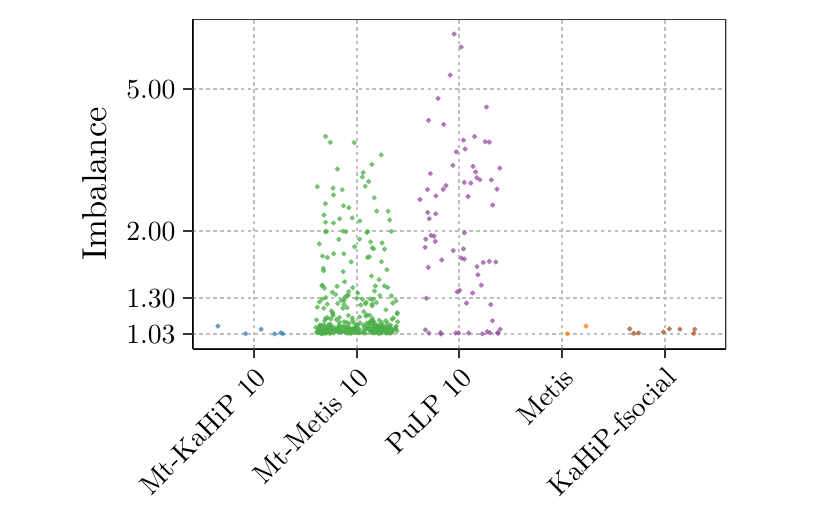}
\fi

      \end{subfigure}
      \begin{subfigure}[t]{0.5\textwidth}
        \vspace*{0em}
        \hspace*{-1.9em}
        
\ifextfig
  \input{tr_plots/largek_shm_imbalance}
\else
  \includegraphics{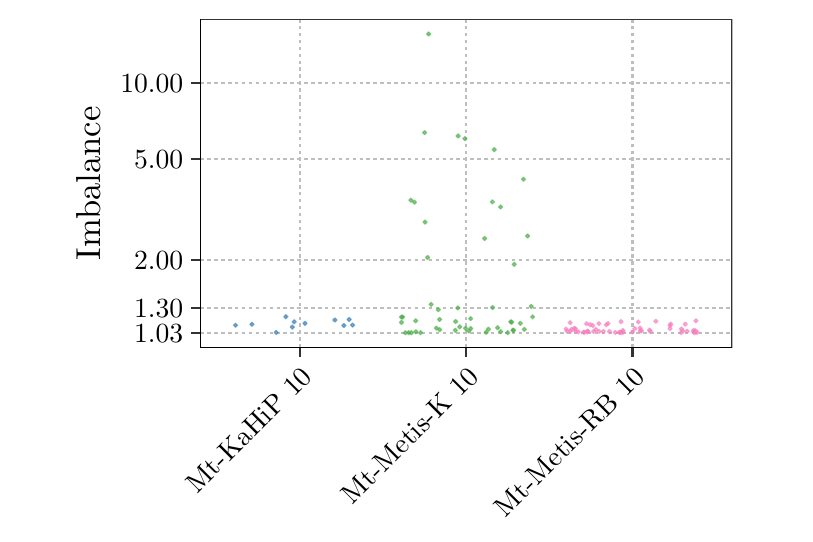}
\fi

      \end{subfigure}
      \vspace*{-1.5em}
      \caption{Left: imbalance of infeasible partitions computed by \Partitioner{Mt-KaHiP 10}, \Partitioner{Mt-Metis 10},
        \Partitioner{PuLP 10}, \Partitioner{Metis} and \Partitioner{KaHiP-fsocial} on benchmark set \textsf{A} with
        $k \in \{2, 4, 8, 16, 32, 64\}$ and $\varepsilon = 0.03$.
        Right: imbalance of infeasible partitions computed by \Partitioner{Mt-Metis-K 10}, \Partitioner{Mt-Metis-RB 10}
        and \Partitioner{Mt-KaHiP 10} on benchmark set \textsf{B} with $k \in \{2^{11}, 2^{14}, 2^{17}, 2^{20}\}$ and $\varepsilon = 0.03$.}
      \label{fig:allk_imbalance}
    \end{figure}
  \end{appendix}
\end{document}